\documentclass[a4paper,floatfix,nofootinbib]{revtex4-2}

\usepackage{amsmath}
\usepackage{amssymb}
\usepackage{bm}
\usepackage{booktabs}
\usepackage{graphicx}
\usepackage{hyperref}

\newcommand{\E}[1]{\ensuremath{\mathbb{E} \! \left[ #1 \right]}}
\newcommand{\Esq}[1]{\ensuremath{\mathbb{E}^2 \! \left[ #1 \right]}}

\newcommand{\K}[1]{\ensuremath{\mathbb{K} \! \left[ #1 \right]}}

\newcommand{\Oof}[1]{\ensuremath{\mathcal{O} \! \left( #1 \right)}}
\newcommand{\Skew}[1]{\ensuremath{\mathbb{S} \! \left[ #1 \right]}}
\newcommand{\Var}[1]{\ensuremath{\mathbb{V} \! \left[ #1 \right]}}
\newcommand{\Vartha}[1]{\ensuremath{\mathbb{V}^{3/2} \! \left[ #1 \right]}}
\newcommand{\vect}[1]{\ensuremath{\mathbf{#1}}}

\begin{document}

\title{A 4\% withdrawal rate for American retirement spending, derived from a discrete-time model of stochastic returns on assets and their sample moments}
\author{Drew M Thomas}
\noaffiliation
\date{7 December 2025}  

\begin{abstract}
What grounds the rule of thumb that a(n American) retiree can safely withdraw 4\% of their initial retirement wealth in their first year of retirement, then increase that rate of consumption with inflation?
I address that question with a discrete-time model of returns to a retirement portfolio consumed at a rate that grows by $s$ per period.
The model's key parameter is $\gamma$, an $s$-adjusted rate of return to wealth, derived from the first 2--4 moments of the portfolio's probability distribution of returns; for a retirement lasting $t$ periods the model recommends a rate of consumption of $\gamma / (1 - (1 - \gamma)^t)$.
Estimation of $\gamma$, and hence of the implied rate of spending in retirement, reveals that the 4\% rule emerges from adjusting high expected rates of return down for: consumption growth, the variance in (and kurtosis of) returns to wealth, the longevity risk of a retiree potentially underestimating $t$, and the inclusion of bonds in retirement portfolios without leverage.
The model supports leverage of retirement portfolios dominated by the S\&P 500, with leverage ratios $> 1.6$ having been historically optimal under the model's approximations.
Historical simulations of 30-year retirements suggest that the model proposes withdrawal rates having roughly even odds of success, that leverage greatly improves those odds for stocks-heavy portfolios, and that investing on margin could have allowed safe withdrawal rates $> 6\%$ per year.
\end{abstract}

\maketitle

\section{Introduction}

The core problem of retirement planning can be expressed as two questions that are inverses of each other.
How much should one save for retirement?
At what rate should one consume one's retirement savings during retirement?
The first question asks, in essence, how much wealth $W$ is necessary to sustain a rate of consumption of $c$ per time period (month, quarter, year) of retirement.
The second question asks, in essence, how high consumption per retirement period $c$ can be for a given $W$.
All else equal, retirement wealth and retirement spending ought to be proportional to each other --- having twice as much in retirement savings can be expected to enable twice as much consumption during retirement --- and so I conclude that $W \propto c$ and that the two questions differ only in which of the two variables they treat as the unknown.
On that assumption, solutions to the core retirement-planning problem amount to estimates of the constant of proportionality between $W$ and $c$.
That is, solving the problem amounts to proposing values of $W/c$ (how much to save for retirement relative to some desired rate of consumption in retirement) or $c/W$ (how quickly one can spend down a given initial stock of retirement wealth).

30 years ago, financial planner William P Bengen ventured a solution to the core retirement-planning problem by analyzing twentieth-century data on returns to investments in US Treasury notes and common stocks \cite{Bengen94}.
He proposed ``safe'' withdrawal rates: estimates of the proportion of saved assets that a retiree could safely liquidate and spend in each year of retirement, increasing their spending in line with inflation over time.
Bengen inferred that, as a rule of thumb, ``a first-year withdrawal of 4 percent [\ldots] should be safe'' for a 30-year retirement \cite[p.\ 173]{Bengen94}.
In other words, $c/W = 4\%$ in the first year of retirement.

How did this number emerge from the probability distribution of rates of return on invested retirement wealth?
Bengen obtained his now-popular 4\% rule of thumb by calculating empirically how actual retirement portfolios, split between US stocks and US bonds, would have performed over 18--50-year retirements starting in various years between 1926 and 1976.
Those direct calculations ensured a degree of realism, but left unclear how rates of return determine in general the maximum ``safe'' $c/W$ or withdrawal rate.
For example, Bengen observed that ``a portfolio consisting of 60-percent stocks and 40-percent bonds could expect an average compounded return of 8.2 percent, assuming continual rebalancing. The `real' return, adjusted for inflation, would be almost 5.1 percent'' \cite[p.\ 172]{Bengen94}.
What, then, would have made a $c/W$ of 5\% unsafe, given a real rate of return just above 5\%?
With 5\% unsafe, why did a $c/W$ of 4\% prove safe?
If historical returns had shown twice as much variability, how much would that have affected the maximum safe $c/W$?

To answer questions like these, which probe the relationship between typical rates of return and the maximum safe $c/W$, I develop a simple model of consumption or ``decumulation'' of wealth during retirement, treating wealth's rate of return as a random variable, and solve the model to obtain approximate but closed-form solutions to the core retirement problem.
Specifically, I derive an approximation of $\E{W/c}$, the expected value of $W/c$, which I invert to obtain an expression for the (initial) withdrawal rate $c/W$.

\label{sectionend:intro}

\section{Model and its approximate solutions}

\subsection{The model}

$W$ represents the initial wealth saved to fund a retirement that lasts for $t$ time periods.
$c$ represents the wealth consumed in the first period of retirement.
Consumption and unconsumed wealth both grow over time.
Unconsumed wealth grows stochastically; in other words, I explicitly model the rate of return to wealth as randomly fluctuating across time periods, and represent that rate of return in period $i$ as the random variable $\vect{r}_i$, where each realization of $\vect{r}_i$ is the relative increase in wealth --- so I am taking ``arithmetic'' returns as the foundation of the model, not ``geometric'' returns.
For simplicity and tractability, I assume that consumption grows at a fixed exponential rate (a rate which can be considered a constant rate of inflation if the retiree's consumption basket never changes).
As such I write the relative growth in consumption per time period as the constant $s$, and consumption in time period $i$ is therefore $c (1+s)^{i-1}$.

A successful retirement in this model is a retirement that consumes all of the initial lump of wealth $W$, and \emph{only} $W$, after exactly $t$ time periods.
That occurs if and only if, after $t$ periods, the wealth lump's counterfactual size under growth without consumption equals the total wealth actually foregone due to consumption:
\begin{equation}
\left( \prod_{i=1}^t 1 + \vect{r}_i \right) W
= \sum_{i=0}^{t-1} (1+s)^i \left( \prod_{j=i+1}^t 1 + \vect{r}_j \right) c
\end{equation}
Rearrangements of this equation can produce an equation for $c/W$ and an equation for $W/c$, either of which one might try to solve.
My approach is to rearrange for $W/c$, not $c/W$, despite $c/W$ being the ultimate target of interest.
I choose that approach because I wish to take an expected value later, and both $W/c$ and $c/W$ are positive and so a priori liable to be positively-skewed when treated as random variables.
The expected value of either ratio would then be influenced quite heavily by the upper tail of the ratio's distribution, so $\E{c/W}$ would suggest an optimistically high rate of consumption spending (the upper tail of $c/W$'s distribution inflating its average) whereas $\E{W/c}$ would suggest conservatively low rates of consumption spending (due to $W/c$'s distribution's upper tail inflating the average amount of wealth needed to fund a given rate of consumption $c$).
I would prefer to arrive at a conservative formula for $c/W$ rather than an optimistic one, and so I should obtain a formula indirectly by inverting $\E{W/c}$.

Deciding to rearrange for $W/c$, I divide through by $c$ and the leftmost product to get
\begin{equation}
\frac{W}{c}
= \sum_{i=0}^{t-1} \; (1+s)^i \! \left( \prod_{j=i+1}^t 1 + \vect{r}_j \right) \Bigg/
\prod_{j=1}^t 1 + \vect{r}_j
\end{equation}
and, cancelling product terms within the sum,
\begin{equation}
\frac{W}{c}
= \sum_{i=0}^{t-1} \frac{ (1+s)^i }{ \prod_{j=1}^i 1 + \vect{r}_j }
\end{equation}
Were $\vect{r}$ not a random variable but a constant, the right-hand side would amount to a geometric sum and this equation would simplify tidily into an explicit formula, but the returns $\vect{r}$ are in fact random; I can suppose only that their general statistical properties might be known.
Therefore, to make progress, I take the expectation of both sides:
\begin{equation}
\E{ \frac{W}{c} }
= \E{ \sum_{i=0}^{t-1} \frac{ (1+s)^i }{ \prod_{j=1}^i 1 + \vect{r}_j } }
\equiv \sum_{i=0}^{t-1} (1+s)^i \, \E{ \prod_{j=1}^i \frac{1}{1 + \vect{r}_j} }
\end{equation}
The rightmost expected value looks troublesome, being an expected value of a rolling product of the reciprocals of the random time series $1 + \vect{r}_j$.
To bring it under control, I normalize the $1 + \vect{r}_j$ terms by first assuming that $\vect{r}$ is a \emph{stationary} process, which entails that $\vect{r}$'s moments, including the expected value, are constant in all time periods.
Then $\E{\vect{r}_j} = \E{r}$ in every time period regardless of $j$, $r$ denoting the return in a single arbitrary period, and I may expand the ``1'' in the numerator of the expectation into $(1 + \E{r}) \div (1 + \E{\vect{r}_j})$ and write:
\begin{equation}
\E{ \frac{W}{c} }
= \sum_{i=0}^{t-1} \frac{ (1+s)^i }{ (1 + \E{r})^i } \E{ \prod_{j=1}^i \frac{1 + \E{\vect{r}_j}}{1 + \vect{r}_j} }
\end{equation}
Assuming next that returns are not only stationary but independently distributed (i.e.\ that the return to wealth in period $j$ provides no information about the return to wealth in any other period), the right-hand side's expectation of the product equals the product of expectations:
\begin{equation}
\E{ \frac{W}{c} }
= \sum_{i=0}^{t-1} \frac{ (1+s)^i }{ (1 + \E{r})^i } \prod_{j=1}^i \E{ \frac{1 + \E{\vect{r}_j}}{1 + \vect{r}_j} }
\label{eq:readyforexpansion}
\end{equation}
Expectations of ratios have no tidy expression, so my tactic at this point is to approximate the ratio as a product, by applying standard Taylor expansions in $X$ of $(1 + X)^{-1}$.
Then, having expanded the expected value of the fraction in powers of $\vect{r}_j$, I truncate the expansion after a few terms to produce a closed-form formula.

\subsection{Second-order approximation and initial results}

Applying the expansion to second order about $\vect{r}_j = 0$,
\begin{align}
\E{ \frac{1 + \E{\vect{r}_j}}{1 + \vect{r}_j} }
& = \E{ \left( 1 + \E{\vect{r}_j} \right)
\left( 1 - \vect{r}_j + \vect{r}_j^2 - \Oof{\vect{r}_j^3} \right) } \notag \\
& \equiv \mathbb{E} \big[
1 + \E{\vect{r}_j} - \vect{r}_j - \E{\vect{r}_j} \vect{r}_j + \vect{r}_j^2 \notag + \E{\vect{r}_j} \vect{r}_j^2 - \Oof{\vect{r}_j^3} \big] \\
& \equiv 1 + \E{\vect{r}_j^2} - \Esq{\vect{r}_j}
+ \E{ \E{\vect{r}_j} \vect{r}_j^2 - \Oof{ \vect{r}_j^3 } }
\end{align}
With $\vect{r}_j \ll 1$,
\begin{equation}
\E{ \frac{1 + \E{\vect{r}_j}}{1 + \vect{r}_j} }
\approx 1 + \E{\vect{r}_j^2} - \Esq{\vect{r}_j}
\equiv 1 + \Var{\vect{r}_j}
\end{equation}
from the general definition of the variance $\Var{X}$ of $X$ as $\E{X^2} - \Esq{X}$.
Thus, to second order, $W/c$'s expectation is
\begin{equation}
\E{ \frac{W}{c} }
\approx \sum_{i=0}^{t-1} \frac{ (1+s)^i }{ (1 + \E{r})^i } \prod_{j=1}^i 1 + \Var{\vect{r}_j}
\end{equation}
From the assumption of stationary returns, the variance $\Var{\vect{r}_j}$ is constant, being $\Var{r}$ in every time period, so
\begin{equation}
\E{ \frac{W}{c} }
\approx \sum_{i=0}^{t-1} \left(
	\frac{ (1+s) (1 + \Var{r}) }{ 1 + \E{r} }
\right)^i
\end{equation}
a geometric sum having the closed-form solution
\begin{equation}
\E{ \frac{W}{c} } \approx \frac{1 - (1 - \gamma)^t}{\gamma}
\textnormal{\ \ where\;}
\gamma
\equiv \frac{ 1 + \E{r} - (1+s)(1 + \Var{r}) }{1 + \E{r}}
\equiv \frac{ \E{r} - (s + \Var{r} + s \Var{r}) }{1 + \E{r}}
\label{eq:secondordersoln}
\end{equation}
In this context $\gamma$ is interpretable as approximately the expected return to wealth adjusted downwards for consumption growth $s$ and variance $\Var{r}$ in returns to wealth.
$\gamma$ can be negative if wealth is expected to produce returns that fail to keep pace with variance and consumption growth.
The interchangeability of $s$ and $\Var{r}$ in the formula indicates that variance in returns behaves exactly (in this second-order approximation) like consumption growth or inflation.

I obtain a formula for the first-period withdrawal rate $c/W$ as the reciprocal of my closed-form $\E{W/c}$ solution:
\begin{equation}
\frac{c}{W}
\approx \left( \E{ \frac{W}{c} } \right)^{-1}
\approx \frac{\gamma}{1 - (1 - \gamma)^t}
\label{eq:secondordercoW}
\end{equation}

Generating a specific number with this formula calls for specific values for $t$ and $\gamma$, and hence numeric estimates of $\E{r}$, $\Var{r}$, and $s$.
The expectation and $s$ are available from Bengen: his ``average compounded return of 8.2 percent'' per year and real return of 5.1\% per year imply $\E{r} = 8.2\%$ and $s = 2.9\%$.
Neglecting $\Var{r}$, $\gamma$ is then 0.049 and, with $t$ = 30, the proposed $c/W \approx 6.3\%$.
This is a significantly more-aggressive withdrawal rate than Bengen's suggested 4\%, and it is unduly aggressive because it relies on assuming away variance in returns.
Supposing more realistically that the variance in returns is similar to consumption's growth rate of 2.9\%, $\gamma$ falls to 0.021 and $c/W$ to 4.5\%, bringing $c/W$ broadly into line with Bengen's 4\% rule of thumb.

With a $c/W$ formula and its derivation I can suggest provisional answers to questions like those posed near the end of section \ref{sectionend:intro}.
A $c/W$ of 5\% was unsafe because it neglected the drag factor of variability in returns;
a $c/W$ of 4\% was safe because, I infer, variability was not great enough to cancel out much of the 5\% real rate of return;
and doubling the variability in historical returns (from the assumed 2.9\%) would reduce $\gamma$ to about $-0.0046$ and hence $c/W$ to about 3.1\%.
Granted, these numbers are speculative because my estimate of $\Var{r}$ is speculative.
Below I re-consider the question of safe annual withdrawal rates given the actual $\Var{r}$ (and higher-order moments of the distribution) of historical time series of returns.

\subsection{Limiting cases of the $c/W$ formula}

Equation \ref{eq:secondordercoW}, the second-order conservative approximate formula for $c/W$, has some simple limits.
An important limit is the $t \rightarrow +\infty$ limit, representing a retirement that lasts forever.
While this is an unrealistic limit, analysis of it accounts for ``longevity risk'' \cite{DeNardi16}, the risk of running out of savings in retirement due to living longer than expected.
Retiring forever should in principle be financially possible if wealth generates positive returns and retirement consumption is sufficiently modest, since the returns to wealth could replenish the wealth spent on consumption.
Equation \ref{eq:secondordercoW}'s $t \rightarrow +\infty$ limit of  supports this intuition mathematically. If $0 < \gamma < 2$ (i.e.\ if returns are positive after adjusting for consumption growth and inflation, but not absurdly positive) then $(1 - \gamma)^t \rightarrow 0$ as $t \rightarrow +\infty$, so
\begin{equation}
\lim_{t \rightarrow +\infty} \frac{c}{W}
\, \approx \, \lim_{t \rightarrow +\infty} \frac{\gamma}{1 - (1 - \gamma)^t}
= \gamma
\end{equation}
revealing that $\gamma$ is interpretable as the (initial) fraction of wealth a retiree can consume per period of an infinitely long retirement.
In other words, $\gamma$ itself turns out to be a conservative estimate of the rate at which a retiree can consume ther initial wealth $W$.

Comparing $\gamma$ to the right-hand side of equation \ref{eq:secondordercoW} suggests an upper bound on the potential impact of longevity risk on the feasible rate of consumption $c/W$.
Dividing $\gamma$ by equation \ref{eq:secondordercoW}'s right-hand side gives the ratio of the feasible $c/W$ for an infinite-duration retirement to the feasible $c/W$ for a retirement lasting only $t$ periods: $1 - (1 - \gamma)^t$.
Switching from a retirement of $t$ periods to an infinitely long retirement therefore cuts the feasible $c/W$ by a fraction of $(1 - \gamma)^t$.
With Bengen's $t$ of 30 years, and my Bengen-inspired $\gamma$ of 0.021 from above, this fraction is 53\%; in the worst case, longevity risk might lead a retiree at age 60 (for example) expecting to live to age 90 to halve their initial retirement spending.

Another limiting case of interest is that of $t$ being finite, not infinite, and $\gamma$ tending to zero.
This corresponds to consumption growth and variance in returns essentially cancelling out the expected return to wealth.
The starkest example of such would be constant retirement consumption ($s = 0$) funded exclusively by cash, defined here as an asset that returns nothing in every period ($\E{r}$ = $\Var{r}$ = 0).
Under constant consumption, as a portfolio becomes all cash, $\gamma \rightarrow 0$ and, for finite $t$, $\gamma t \rightarrow 0$.
Then, applying the binomial expansion of $(1 - \gamma)^t$ to first order,
\begin{equation}
\lim_{\gamma \rightarrow 0} \frac{c}{W}
\, \approx \, \lim_{\gamma \rightarrow 0} \frac{\gamma}{1 - (1 - \gamma)^t}
\, \rightarrow \, \lim_{\gamma \rightarrow 0} \frac{\gamma}{1 - (1 - \gamma t)}
= \frac{1}{t}
\end{equation}
implying that such a retirement lasting $t$ periods means consuming $1/t$ of the initial wealth per period.
This result is intuitively sensible.
If returns to wealth approximately compensate for consumption growth (and variance in returns), that fact is much more important than the precise gap between returns and consumption growth; that is, the precise value of $\gamma$ ceases to matter and the duration of retirement becomes the key determinant of the feasible spending rate.
In each year of a 30-year retirement, for example, the retiree would simply spend $1/30$ of the wealth with which they began retirement.

\subsection{Optimal leverage as limited leverage}

Augmenting the formula for $\gamma$ can also explain why trying to arbitrarily boost a portfolio's expected return by indefinitely levering it up is a poor retirement-funding tactic.
Introducing leverage as the leverage ratio $l$, $\gamma$ becomes, in the second-order approximation,
\begin{equation}
\gamma
\equiv \frac{ 1 + \E{lr} - (1+s)(1 + \Var{lr}) }{1 + \E{lr}}
\equiv 1 - (1+s) \frac{1 + l^2 \Var{r}}{1 + l \E{r}}
\end{equation}
since leverage serves as a multiplier on returns; until this point I have effectively assumed $l=1$, an un-levered portfolio.

A higher $\gamma$ is desirable because $\gamma$ is, crudely, an adjusted expected return to wealth, and so the formula immediately suggests why sufficiently aggressive leverage would be troublesome.
Using leverage multiplies $\E{r}$ by $l$ but multiplies $\Var{r}$ by the \emph{square} of $l$, so sufficiently large leverage grows the rightmost fraction's numerator more than its denominator, hence enlarging the fraction as a whole and reducing $\gamma$.
Were $\Var{r}$ literally nil, multiplying $\E{r}$ by an arbitrarily large $l$ could only increase $\gamma$, but for any positive $\Var{r}$ there is always an $l$ large enough that the drag of variance dominates and causes $\gamma$ to decline with further leverage.

The specific optimal $l$ is the $l$ that maximizes $\gamma$, which I now deduce.
The first derivative of $\gamma$ with respect to $l$ is
\begin{equation}
\frac{\partial \gamma}{\partial l}
= - (1+s) \frac{ (1 + l \E{r}) (2l \Var{r}) - \E{r} (1 + l^2 \Var{r})  }{ (1 + l \E{r})^2 }
\end{equation}
and this derivative is zero when $\gamma$ is maximized.
(Theoretically this derivative would also be zero whenever $\gamma$ were \emph{min}imized, too, but the \emph{second} derivative is negative under the assumptions of the next sentence, so on those assumptions a zero first derivative here represents a maximum of $\gamma$.)
Assuming $s > -1$, $l$ non-negative and finite, and $\E{r}$ non-negative and finite, the first derivative is zero when the numerator of its fraction is zero, i.e.\ when
\begin{equation}
(1 + l \E{r}) (2 l \Var{r}) - \E{r} (1 + l^2 \Var{r}) = 0
\quad \Longleftrightarrow \quad
\E{r} \Var{r} l^2 + 2 \Var{r} l - \E{r} = 0
\end{equation}
a quadratic equation which has the solutions
\begin{equation}
l = \frac{-2 \Var{r} \pm \sqrt{
	4 \mathbb{V}^2 [r] + 4 \Esq{r} \Var{r}
}}{2 \E{r} \Var{r}}
\equiv
\frac{- \Var{r} \pm \sqrt{
	\mathbb{V}^2 [r] + \Esq{r} \Var{r}
}}{\E{r} \Var{r}}
\end{equation}
In general the two solutions are of different sign.
When $\E{r} > 0$, the positive-$l$ solution is
\begin{equation}
l = \frac{\sqrt{ \mathbb{V}^2 [r] + \Esq{r} \Var{r} } - \Var{r}}{\E{r} \Var{r}}
\equiv \frac{\sqrt{ 1 + \frac{\Esq{r}}{\Var{r}} } - 1}{\E{r}}
\label{eq:secondorderleverageratio}
\end{equation}
and this is, in the second-order approximation, the feasible leverage ratio that maximizes $\gamma$.

With my loosely Bengen-inspired estimates of $\E{r} =$ 8.2\% and $\Var{r} =$ 2.9\% for annual returns to a Bengen portfolio (60\% ``common stocks'' and 40\% ``intermediate-term Treasuries''), the second-order-optimal leverage ratio $l$ is 1.34: a retiree with access to interest-free margin loans could consume more by borrowing an additional third of their initial retirement wealth at the beginning of retirement to buy extra (US) stocks and (US Treasury) bonds.
Above I calculate a $\gamma$ of 0.021 for an un-levered ($l=1$) Bengen portfolio; with $l = 1.34$, $\gamma$ rises to 0.025, and the first-year withdrawal ratio $c/W$ for a 30-year retirement rises accordingly from 4.5\% to 4.7\%.

The finding that leverage can enhance the portfolio's ability to sustain a retirement may be uninteresting, especially given the unrealistic assumption of free leverage.
Of more interest may be the model's ability to highlight why leverage should be finite even if leverage is free: leverage amplifies variance in returns more than expected returns.

\subsection{Fourth-order approximation}

The second-order approximation demonstrates my basic analytical strategy and produces a crude relationship between a reasonable withdrawal rate and the period-to-period variability of returns, but it is quite imperfect, using only the first two moments of the probability distribution of returns.
An obvious improvement is to incorporate higher-order moments of the returns distribution by expanding the rightmost fraction to higher order.
I now carry out that improvement by expanding the fraction to fourth order, incorporating the skewness and kurtosis of returns as well as returns' mean and variance, before calculating how that alters the practical results.

I simplify my notation for brevity, writing $\vect{r}_j$ as just $r$ (licensed by my earlier assumption of stationarity).
I also change how I carry out the expansion, expanding eq.\ \ref{eq:readyforexpansion}'s relevant fraction about $r = \E{r}$ instead of expanding about $r = 0$.
Expanding about $r$'s mean instead of zero produces an expansion in terms of returns' central moments instead of returns' raw moments, a more-intuitive expansion and one that requires less tedious manipulation to be re-expressed in terms of standardized moments.
The relevant fraction's expansion about $r = \E{r}$ is
\begin{equation}
\frac{1 + \E{r}}{1 + r}
= \sum_{p=0}^{\infty} \, (-1)^p \! \left( \frac{r - \E{r}}{1 + \E{r}} \right)^p
\end{equation}
Writing it out to fourth order and taking its expectation term by term,
\begin{equation}
\E{ \frac{1 + \E{r}}{1 + r} }
= 1
- \frac{ \E{r - \E{r}} }{1 + \E{r}}
+ \frac{ \E{(r - \E{r})^2} }{(1 + \E{r})^2}
- \frac{ \E{(r - \E{r})^3} }{(1 + \E{r})^3}
+ \frac{ \E{(r - \E{r})^4} }{(1 + \E{r})^4}
- \E{ \Oof{ \left( \frac{r - \E{r}}{1 + \E{r}} \right)^5 } }
\end{equation}
The 5 explicit terms reflect $r$'s zeroth through fourth central moments.
Rewriting them in terms of $r$'s standardized moments,
\begin{equation}
\E{ \frac{1 + \E{r}}{1 + r} }
= 1
+ \frac{ \Var{r} }{(1 + \E{r})^2}
- \frac{ \Vartha{r} \Skew{r} }{(1 + \E{r})^3}
+ \frac{ \mathbb{V}^2[r] \K{r} }{(1 + \E{r})^4}
- \E{ \Oof{ \left( \frac{r - \E{r}}{1 + \E{r}} \right)^5 } }
\end{equation}
where the first standardized moment, being zero, disappears entirely, and $\Skew{r}$ and $\K{r}$ denote the third and fourth standardized moments respectively.
Truncating the series after the fourth-order/fourth-moment term, and factoring out the second-order multiplier common to the third- and fourth-order terms,
\begin{equation}
\E{ \frac{1 + \E{r}}{1 + r} }
\approx 1
+ \frac{ \Var{r} }{(1 + \E{r})^2} \left( 1
- \frac{ \mathbb{V}^{1/2}[r] \Skew{r} }{1 + \E{r}}
+ \frac{ \mathbb{V}[r] \K{r} }{(1 + \E{r})^2}
\right)
\end{equation}
Substituting into my earlier expression for $\E{W/c}$,
\begin{equation}
\E{ \frac{W}{c} }
\approx \sum_{i=0}^{t-1} \left(
	\frac{1+s}{ 1 + \E{r} }
	\left( 1 + \frac{ \Var{r} }{(1 + \E{r})^2}
		\left( 1
			- \frac{ \mathbb{V}^{1/2}[r] \Skew{r} }{1 + \E{r}}
			+ \frac{ \Var{r} \K{r} }{(1 + \E{r})^2}
		\right)
	\right)
\right)^i
\end{equation}
a geometric sum of essentially the same form as before, and having a solution of essentially the same form ---
\begin{equation}
\E{ \frac{W}{c} } \approx \frac{1 - (1 - \gamma)^t}{\gamma}
\quad \Longrightarrow \quad
\frac{c}{W} \approx \frac{\gamma}{1 - (1 - \gamma)^t}
\end{equation}
--- but now with a more-complicated $\gamma$ that accounts for two more moments of the returns distribution:
\begin{equation}
\gamma
\equiv 1 - \frac{1+s}{ 1 + \E{r} } \left(
	1 + \frac{ \Var{r} }{ ( 1 + \E{r} )^2 } \left( 1
		- \frac{ \mathbb{V}^{1/2}[r] \Skew{r} }{1 + \E{r}}
		+ \frac{ \mathbb{V}[r] \K{r} }{(1 + \E{r})^2}
	\right)
\right)
\label{eq:fourthordersoln}
\end{equation}
There is an interesting second-order difference between this result and my earlier second-order $\gamma$ formula (eq.\ \ref{eq:secondordersoln}).
Unlike the earlier formula, this formula indicates that the leading factor of $\Var{r}$ should be divided by $(1 + \E{r})^2$, implying that the earlier formula pessimistically overstates the impact of $\Var{r}$ when $\Var{r}$ is very small.
However, some comparisons with rough simulation results suggest that the overstatement is usefully conservative, because second-order formulae for $\gamma$ understate the impact of $\Var{r}$ when $\Var{r}$ is larger, and the earlier formula's pessimistic overstatement partially compensates for that optimistic large-$\Var{r}$ bias.

In this higher-order $\gamma$ formula, $\Var{r}$ appears exclusively divided by $(1 + \E{r})^2$, so one may write the formula as
\begin{equation}
\gamma
\equiv 1 - \frac{1+s}{ 1 + \E{r} } \left(
	1 + \tilde{\sigma}^2 \left( 1
		- \tilde{\sigma} \Skew{r}
		+ \tilde{\sigma}^2 \K{r}
	\right)
\right)
\textnormal{\ \ where\;}
\tilde{\sigma} \equiv \frac{ \mathbb{V}^{1/2}[r] }{ 1 + \E{r} }
\end{equation}
and $\tilde{\sigma}$ might be considered a reduced standard deviation of returns, or a coefficient of variation of $1 + \vect{r}$.

However it is represented mathematically, the fourth-order analysis can systematically underestimate $\E{W/c}$ and hence overestimate $\gamma$.
This becomes clear when considering the infinite-series expansion taken about $r = \E{r}$ if $r$ has a symmetric probability distribution.
In that case, the expectations of odd-numbered powers of $(r - \E{r})$ vanish, leaving only the sum of expectations of even-numbered powers, each of which must be positive if $\Var{r} > 0$.
The truncation of the infinite series at \emph{any} power or order therefore induces an underestimate of $\E{W/c}$ for symmetric distributions, since the truncation omits the positive contribution of higher-order even-numbered powers/moments.

Bengen's original paper provides only the first moment of the returns distribution, so assumptions about the next 3 moments are necessary to try to compare my fourth-order formula's results to Bengen's.
Supplementing my earlier assumption that $\Var{r} = s = 2.9$\% with the (unrealistic) assumption that the distribution of returns is symmetric ($\Skew{r}$ = 0) and mesokurtic ($\K{r} = 3$), the fourth-order formula produces $\gamma = 0.021$, the same to within rounding error as my second-order estimate.
However, the effect of kurtosis need not always be negligible; a higher (but not absurd) kurtosis of 5 instead of 3 would reduce $\gamma$ to 0.019.
Not only does returns' variance drag down the rate of consumption of a retirement portfolio, but so does returns' kurtosis.

\section{Application of model-derived formulae}

\subsection{Actual returns' moments and implied withdrawal rates}

To apply the fourth-order formulae to more-complete estimates of the higher-order moments of probability distributions of returns, I move on to estimating the moments of those distributions myself.

My first step in this direction was to download monthly time series of the returns of exchange-traded funds (ETFs) from Yahoo!\ Finance in spring 2024 (before Yahoo!\ Finance eliminated public access to historical asset prices).
I selected the oldest ETFs I could find that represented the S\&P 500 index of US equities, and US Treasury Bonds of different durations.

Actual ETF prices offer the advantage of reflecting the actual fees charged to hold the relevant collections of assets.
However, the relevant ETFs are relatively young ($< 35$ years) with accordingly short time series of returns.
I therefore pay more attention to the much longer, albeit coarser-grained, time series of Shiller, which extend back through 1871.
This is not a trivial methodological decision.
Comparing Shiller's returns time series to the ETF returns time series (table \ref{ta:etfsindices}) indicates that returns of both the S\&P 500 and Treasury Bonds systematically differed in recent years versus earlier years: equities have exhibited higher recent returns (albeit with more variability and negative skewness) and bonds much lower returns (albeit with positive skewness and less kurtosis).

\begin{table*}
\begin{ruledtabular}
\caption{\label{ta:etfsindices} ETFs and indices representing different asset classes: their ticker symbols (applicable only to ETFs), their names and start dates, the first four central/standardized sample moments of their monthly returns $\vect{r}$ (also written as ``$r$''), $\tilde{\sigma}$, and two $\gamma$ estimates, given monthly consumption growth $s$ of 0.3\%.}
\begin{tabular}{ccccrrrrrrrr}
\multicolumn{3}{c}{ETF or index} & & & \multicolumn{4}{c}{moments of monthly $\vect{r}$} & & \multicolumn{2}{c}{$\gamma$ estimates (\%)} \\
\cmidrule{1-3} \cmidrule{6-9} \cmidrule{11-12}
sym.\ & name of ETF/index & start date & $N$ & asset class & $100 \E{r}$ & $10^4 \Var{r} $ & skew.\ & \K{r} & $100 \tilde{\sigma}$ & $\gamma_2$ & $\gamma_4$ \\
\hline
& \footnotesize Shiller's S\&P time series & Jan.\ 1871 & 1850 & US equities
& 0.823 & 16.4 & 0.446 & 20.5 & 4.02 & 0.355 & 0.356 \\
& \footnotesize Shiller's ``GS10'' interest rates & Jan.\ 1871 & 1850 & US bonds
& 0.383 & 1.65 & 1.06 & 14.5 & 1.28 & 0.0663 & 0.0666 \\
\hline
\texttt{SPY} & \footnotesize SPDR S\&P 500 & 1993-01-22 & 375 & US equities
& 0.901 & 19.1 & $-0.520$ & 3.88 & 4.33 & 0.406 & 0.404 \\
\texttt{AGG} & \footnotesize iSh.\ Core U.S.\ Aggregate Bond & 2003-09-22 & 247 & US bonds
& 0.243 & 1.72 & 0.228 & 5.57 & 1.31 & $-0.0743$ & $-0.0742$ \\
\hline
\texttt{SHV} & \footnotesize iSh.\ Short Treasury Bond & 2007-01-05 & 207 & T-bills
& 0.101 & 0.0418 & 3.08 & 16.2 & 0.204 & $-0.199$ & $-0.199$ \\
\texttt{SHY} & \footnotesize iSh.\ 1--3 Year Treasury Bond & 2002-07-22 & 262 & T-notes
& 0.143 & 0.211 & 0.517 & 6.43 & 0.458 & $-0.158$ & $-0.158$ \\
\texttt{IEI} & \footnotesize iSh.\ 3--7 Year Treasury Bond & 2007-01-05 & 208 & T-notes
& 0.219 & 1.36 & 0.112 & 3.71 & 1.16 & $-0.0945$ & $-0.0944$ \\
\texttt{IEF} & \footnotesize iSh.\ 7--10 Year Treasury Bond & 2002-07-22 & 262 & T-notes
& 0.291 & 3.80 & 0.138 & 3.71 & 1.94 & $-0.0470$ & $-0.0467$ \\
\texttt{TLH} & \footnotesize iSh.\ 10--20 Year Treasury Bond & 2007-01-05 & 208 & T-securits.\ 
& 0.275 & 8.78 & 0.336 & 3.96 & 2.96 & $-0.112$ & $-0.111$ \\
\texttt{TLT} & \footnotesize iSh.\ 20+ Year Treasury Bond & 2002-07-22 & 262 & T-bonds
& 0.369 & 15.7 & 0.364 & 4.23 & 3.94 & $-0.0879$ & $-0.0855$ \\
\end{tabular}
\end{ruledtabular}
$N$ is the number of monthly observations in $\vect{r}$.
``iSh.'' in an ETF's name denotes ``iShares''.
\end{table*}

Table \ref{ta:etfsindices} also collects the results of applying two formulae for $\gamma$, which I label ``$\gamma_2$'', the second-order $\gamma$ from equation \ref{eq:secondordersoln}, and ``$\gamma_4$'', the fourth-order $\gamma$ from equation \ref{eq:fourthordersoln}.
This allows an immediate judgement of the impact of the formulae's differences.
For consistency, I use across all $\gamma$ estimates the same value for $s$ of 0.3\%, the rounded mean of monthly US CPI inflation \cite{FREDCPI}.
(A monthly $s$ of 0.3\% annualizes to consumption growth of 3.66\% per year, modestly outpacing historical inflation.)
For each ETF and index, the difference between $\gamma_2$ and $\gamma_4$ is small, suggesting that higher-order moments of the probability distributions of returns are relatively unimportant (though this is partly the coincidence of positive skewnesses and positive excess kurtoses offsetting each other's impact on $\gamma$).

Regardless of formula, $\gamma$ has proven patently negative for US-bond ETFs.
$\gamma$ being negative signals that the average return is insufficient to compensate for consumption growth and the variability in returns.
Illustrating with a concrete example, substituting $t = 12 \times 30$ (representing a 30-year retirement) and $\gamma = -0.000742$ (the fourth-order $\gamma$ for the \texttt{AGG} bond ETF) into eq.\ \ref{eq:secondordercoW} gives a $c/W$ estimate of 0.00242, corresponding to a withdrawal rate of 3.0\% per year, slightly worse than the na\"{i}ve rate of 3.3\% per year one would obtain by dividing 100\% by 30 years.

Large-cap US equities have proven far more useful for overcoming inflation and variance drag.
In recent years the S\&P 500 ETF \texttt{SPY} has shown a strong average return of 0.9\% per month.
As such, despite the variance, negative skewness, and leptokurtic nature of its returns, the 1993--2024 $\gamma_4$ of \texttt{SPY} is 0.00404, producing a $c/W$ of 0.00526 and so a withdrawal rate of 6.4\% in the first year of a 30-year retirement.
I do not propose that seriously as a safe withdrawal rate for a simple all-equities retirement portfolio, but am merely illustrating the power of a high average rate of return.
The 6.4\% rate could equally be read as illustrating the power of consumption growth and variance in returns: even that optimistic rate is notably less than the (annualized) average rate of return of 11.4\% per year.
The withdrawal rate must be significantly less to guard against not just inflation but fluctuations in returns.

One reason not to take seriously the $c/W$ of 0.00526 and implied annual withdrawal rate of 6.4\% is the limited history they represent.
Switching to Shiller's long-run data produces a lower $\gamma_4$ of 0.00356 and $c/W$ of 0.00492, and so a lower annual withdrawal rate of 6.0\%.

This is still notably greater than Bengen's classic 4\%.
Portfolio composition explains much of the remaining gap between the 6.0\% and Bengen's 4\%; Bengen's 4\% came from assuming a portfolio that was half stocks and half bonds.
Under Shiller's time series and monthly rebalancing (rebalancing now being a necessary factor to consider since the portfolio is heterogeneous), $\gamma_4$ falls to 0.00259, $c/W$ to 0.00427, and the annual withdrawal rate to 5.2\%.
Accounting for outperformance of stocks since Bengen's article reduces the numbers further: using Shiller's returns only through December 1992 reduces $\gamma_4$ slightly to 0.00248, $c/W$ to 0.00420, and the annual withdrawal rate to 5.1\%.

The remaining gap between my annual withdrawal rate and Bengen's 4\% comes from a fundamental conceptual difference between my approach and Bengen's: my algebra is based on deriving \emph{expectations} (that is, averages), whereas Bengen accounts for something close to a worst-case scenario.
In fact, figure 1(c) of Bengen's paper, which plots how long a 50/50 portfolio would have lasted given different start years and a 5\% annual withdrawal rate, suggests that a typical retirement would indeed have lasted about 30 years, broadly consonant with my own 5.1\% estimate for a 30-year retirement.
(Granted, the consonance is exaggerated somewhat by Bengen capping the effective height of his plot --- the maximum duration of retirement --- at 50 years.)
By contrast, the prose of Bengen's paper indicates that his concern was less with achieving a \emph{typical} retirement duration of 30 years, but more with a ``safe'' retirement \cite[p.\ 173]{Bengen94} that entailed a \emph{minimum} retirement duration of 30 years; Bengen made a point of noting that ``[i]n no past case'' did ``a first-year withdrawal of 4 percent [\ldots] cause[\ldots] a portfolio to be exhausted in 33 years''.
Raising my $t$ from $12 \times 30$ months to $12 \times 33$ months, $c/W$ comes down to 0.00396 and the annual withdrawal rate to 4.8\%.
Rounding down then reproduces Bengen's classic withdrawal rate of 4\%.
Raising $t$ further to $12 \times 46$ months, which appears to be the longest un-capped portfolio lifetime in Bengen's figure 1(b) --- Bengen's graph for a 50/50 portfolio with a 4\% annual withdrawal rate --- lowers my $c/W$ to 0.00332 and the annual withdrawal rate to 4.1\%, aligning my results with Bengen's rule to within rounding error.

Thus my algebra arrives, by a first-principles mathematical route, at a similar destination to Bengen's empirical year-by-year approach.
The S\&P 500 has provided impressive returns of over 10\% per year in the long run, but the four horsemen of (\emph{i}) inflation and consumption growth, (\emph{ii}) variance and kurtosis drags, (\emph{iii}) longevity risk, and (\emph{iv}) the common practice of diluting large-cap US stocks with US bonds, collectively drive the maximum safe annual withdrawal rate far below the headline rate of return: 4\% per year, or a little more, but not 5\% per year.
Retirement advice which accounts for only the first and last of the four factors, like Dave Ramsey's advice to withdraw 8\% per year (based on subtracting 4 percentage points of inflation from a 12\% rate of return) \cite{Bontrager23,Ramsey23}, is flawed.
Such proposals dangerously neglect longevity risk and variance and kurtosis drags.
Far from being ``ridiculous'', ``trash'', ``crap'', and ``stupidity'', lower annual withdrawal rates of 3\%--5\% better reflect \emph{all} of the factors my model incorporates, in the process answering Ramsey's indignant ``if you think you can only pull off 4\% off of investments making 12, where the \emph{flip} is the other 8\% going?!'' query \cite{Bontrager23,Ramsey23}.

\subsection{Optimal leverage redux: borrowing costs as a tighter constraint on leverage}

Equipped with estimates of $\E{r}$ and $\Var{r}$ from Shiller's extensive historical time series, I return to equation \ref{eq:secondorderleverageratio}, my formula for the second-order-optimal leverage ratio $l$.
In a previous section I estimated an optimal $l$ of 1.34 for a 60/40 portfolio, using approximate Bengen-inspired $\E{r}$ and $\Var{r}$ values.
Drawing now on Shiller's time series instead, the 60/40 portfolio has $\E{r} = 0.65\%$ and $\Var{r} = 0.060\%$ for its monthly returns, implying an optimal $l$ of 5.31; for stocks and bonds alone (for which table \ref{ta:etfsindices} directly states $\E{r}$ and $\Var{r}$) the optimal $l$s are 2.48 and 11.39 respectively.

Those leverage ratios are implausibly high, an obvious reason for which is that they neglect the cost of leverage, in effect assuming that leverage is free.
To obtain more-credible estimates of the optimal $l$, I re-derive the formula for the $l$ that maximizes $\gamma$, but this time explicitly include the cost of borrowing.

Introducing the price of leverage as $q$, the interest rate charged on borrowing represented as a proportion of the portfolio per period, the return $r$ in a given period is not only multiplied by $l$ but also reduced by $(l-1)q$, $l-1$ being the borrowed multiple of the portfolio's original un-levered value.
$\gamma$ is then
\begin{equation}
\frac{
	1 + \E{lr - (l-1)q} - (1+s)(1 + \Var{lr - (l-1)q})
}{1 + \E{lr - (l-1)q}}
\equiv 1 - (1+s) \frac{
	1 + l^2 \Var{r} + (l-1)^2 \Var{q} - 2 l (l-1) \textnormal{Cov}[r,q]
}{1 + l \E{r} - (l-1) \E{q}}
\end{equation}
where the original derivation of the optimal $l$ effectively assumed $q=0$.

It is possible to solve this more-complicated $\gamma$ expression for the $l$ that maximizes it, but for concision, I impose the simplifying assumption that $r$ and $q$ are uncorrelated, i.e.\ that $\textnormal{Cov}[r,q] = 0$.
Then
\begin{equation}
\gamma \approx 
1 - (1+s) \frac{
	1 + l^2 \Var{r} + (l-1)^2 \Var{q}
}{1 + l \E{r} - (l-1) \E{q}}
\end{equation}
the derivative of which with respect to $l$ is rather long-winded.
Solving for the $l$ that renders the derivative zero leads, after tedious algebra, to the $\gamma$-maximizing $l$:
\begin{equation}
l = \frac{
	\sqrt{\frac{(1 + \E{q})^2 \Var{r} + (1 + \E{r})^2 \Var{q} + (\E{r} - \E{q})^2}{\Var{r} + \Var{q}}} - (1 + \E{q})
}{\E{r} - \E{q}}
\label{eq:costlyleverageratio}
\end{equation}
if $\E{r} > \E{q}$ (i.e.\ if the portfolio's average return is greater than the average borrowing rate charged to lever it).
Reassuringly, substituting $\E{q} = \Var{q} = 0$ into this formula simplifies it to equation \ref{eq:secondorderleverageratio}.

Taking monthly 3-month T-bill rates \cite{FREDTBill} to represent the price(s) of leverage $q$, $\E{q} = 0.277\%$ and $\Var{q} = 6.13 \times 10^{-6}$.
Using those values, and the entirety of Shiller's 1871--2025 time series, to recompute $l$ for the 3 candidate portfolios considered so far, the second-order-optimal $l$s are 3.05 for the 60/40 portfolio, 1.65 for the all-stocks portfolio, and 3.14 for the all-bonds portfolio (table \ref{ta:leveredportfolios}).
These are all far less than the $l$s obtained under the na\"{i}ve assumption of free leverage, confirming that it is vital to account for the price of leverage if deciding how much leverage to employ.

\begin{table*}
\begin{ruledtabular}
\caption{\label{ta:leveredportfolios} Theoretically derived $\gamma$, first-month withdrawal rate $c/W$, and AWR (annual withdrawal rate) of 30-year retirements funded by an un-levered portfolio ($l = 100\%$) or a second-order-optimally levered ($l > 100\%$) portfolio, assuming leverage priced around the 3-month T-bill rate. All quantities in the main body/matrix of the table are percentages.}
\begin{tabular}{rrrrrrrrrrrrr}
& & & & & \multicolumn{8}{c}{second-order-optimal leverage} \\
\cmidrule{6-13}
\multicolumn{2}{c}{portfolio} & \multicolumn{3}{c}{$l = 100\%$} & \multicolumn{4}{c}{1871--2025 $\E{r}$ and $\Var{r}$} & \multicolumn{4}{c}{1934--2025 $\E{r}$ and $\Var{r}$} \\
\cmidrule{1-2} \cmidrule{3-5} \cmidrule{6-9} \cmidrule{10-13}
equities & bonds & $\gamma$ & $c/W$ & AWR & $l$ & $\gamma$ & $c/W$ & AWR & $l$ & $\gamma$ & $c/W$ & AWR \\
\hline
 60 &  40 & 0.285  & 0.444 & 5.42 & 305 & 0.537 & 0.627 & 7.65 & 438 & 0.956 & 0.987 & 12.0 \\
100 &   0 & 0.355  & 0.492 & 6.00 & 165 & 0.425 & 0.542 & 6.61 & 240 & 0.769 & 0.820 & 10.0 \\
  0 & 100 & 0.0663 & 0.312 & 3.81 & 314 & 0.145 & 0.356 & 4.34 & 233 & 0.129 & 0.347 & 4.23 \\
\end{tabular}
\end{ruledtabular}
The $l = 100\%$ portfolio uses the means and variances of Shiller's entire 1871--2025 time series of bond and equity returns.
\end{table*}

The effect of borrowing costs is likely to be greater still in practice, because I have neglected the correlation between the portfolio's return and the cost of borrowing over time.
Fortunately, the correlation between the latter and most of the assets/indices considered in table \ref{ta:etfsindices} is very small.
The 3-month T-bill rates have their strongest correlations with, unsurprisingly, the returns of T-bills ETF \texttt{SHV} (a Pearson product-moment correlation coefficient of +0.64) and T-notes ETF \texttt{SHY} (a correlation coefficient of +0.21).
All of the other correlations are less than 0.15 in magnitude, small and even negative for long-term Treasury bonds, and negligible (between $-0.03$ and +0.02) for US equities.

Far more relevant to the optimal leverage ratio is the time period over which to evaluate returns.
My estimates of the price of leverage are available only from January 1934, so there is an obvious case for calculating optimal leverage according to equity and bond returns only from 1934 onwards, rather than using the entirety of Shiller's 1871--2025 time series.
The rightmost columns of table \ref{ta:leveredportfolios} collect the results of doing so; the optimal leverage ratios for the mostly-equities portfolios increase by 43--45\%, and the optimal $l$ of the all-bonds portfolio decreases by a quarter.
That is a strong suggestion that no optimal $l$ derived from the model should be taken seriously all of the way to the third significant figure; the results are too sensitive to choices of time period.

Practically-minded analysts may rightly level the further objection that everyday retirees would not have the cheapest access to leverage.
US brokerages generally charge higher consumer margin rates than whatever market lending rate they use as a benchmark.
Their choice of benchmark, moreover, is more likely to be the (effective) federal funds rate (EFFR) \cite{FREDFEDFUNDS} than a T-bill rate, and the EFFR is generally higher than the 3-month T-bill rate I have been using as the price of leverage: in the 1954--2025 period when both rates were available, the EFFR averaged 86 basis points above the 3-month T-bill rate, and was greater than the T-bill rate in 62\% of months.

Accounting for everyday retirees' higher price of leverage does affect the second-order-optimal level of leverage.
For a portfolio worth \$0.1M--\$1M, Robinhood and Interactive Brokers --- the brokerages offering to non-professional US investors the cheapest margin rates I know of --- charge 90--100 basis points per year above the annual EFFR as of mid-2025.
Taking the price of leverage accordingly as the EFFR plus a spread of 100 basis points per year raises $\E{q}$ and $\Var{q}$ to 0.451\% and $7.71 \times 10^{-6}$ respectively.
Applied to data for mid-1954 (when the EFFR time series begins) onwards, the optimal $l$s are 3.09 for the 60/40 portfolio, 1.92 for equities alone, and just 0.10 for bonds alone, all lower than the optimal $l$s derived from 1934--2025 asset prices (especially, of course, for bonds).

My inclination would be to err, for a given portfolio, towards the lowest of the $l$s calculated for that portfolio.
That approach suggests using no leverage if investing exclusively in US Treasury bonds, and using leverage ratios of 1.6--3.1 for monthly-rebalanced portfolios mostly comprising large-cap US stocks.

Nonetheless, using leverage was not only second-order optimal for mostly-equities portfolios, but could have produced meaningful gains, even assuming margin priced at the EFFR plus a percentage point.
For a classic 60/40 portfolio, applying leverage (with a rather aggressive leverage ratio of 309\%, at that) might have enabled a first-year withdrawal of 8.2\% of holdings rather than only 5.4\%.
The optimal leverage for an all-equities portfolio was less, but would still have allowed a first-year withdrawal of 8.3\%, as opposed to 6.0\% for the un-levered retirement portfolio.
These translate to hundreds if not thousands of additional dollars of consumption per year of retirement.

Finally, I note that sufficiently cheap leverage enables the 60/40 portfolio to outperform the all-equities portfolio (table \ref{ta:leveredportfolios}), consistent with Clifford Asness's long-standing prescription to diversify a portfolio and then lever it up as desired, instead of investing entirely in stocks just because their expected return is highest \cite{Asness96}.
Leverage makes possible the full enjoyment of the benefits of diversification, mitigating the drag on returns that lower-return assets might otherwise impose.

\section{Historical performance of the formulae's recommendations}

That my formulae have semi-rigorous mathematical derivations does not guarantee that they are a reliable guide to reality.
I therefore now assess how successful retirements would have been, given historical asset returns, if funded according to my formulae.
In particular, I test my second-order formulae becausee they are cruder (if they prove adequate, then my fourth-order formulae should be adequate too), and because I obtained my optimal-leverage formulae by approximating to only second order.

\begin{figure*}
\centering
\includegraphics[width=0.999\linewidth]{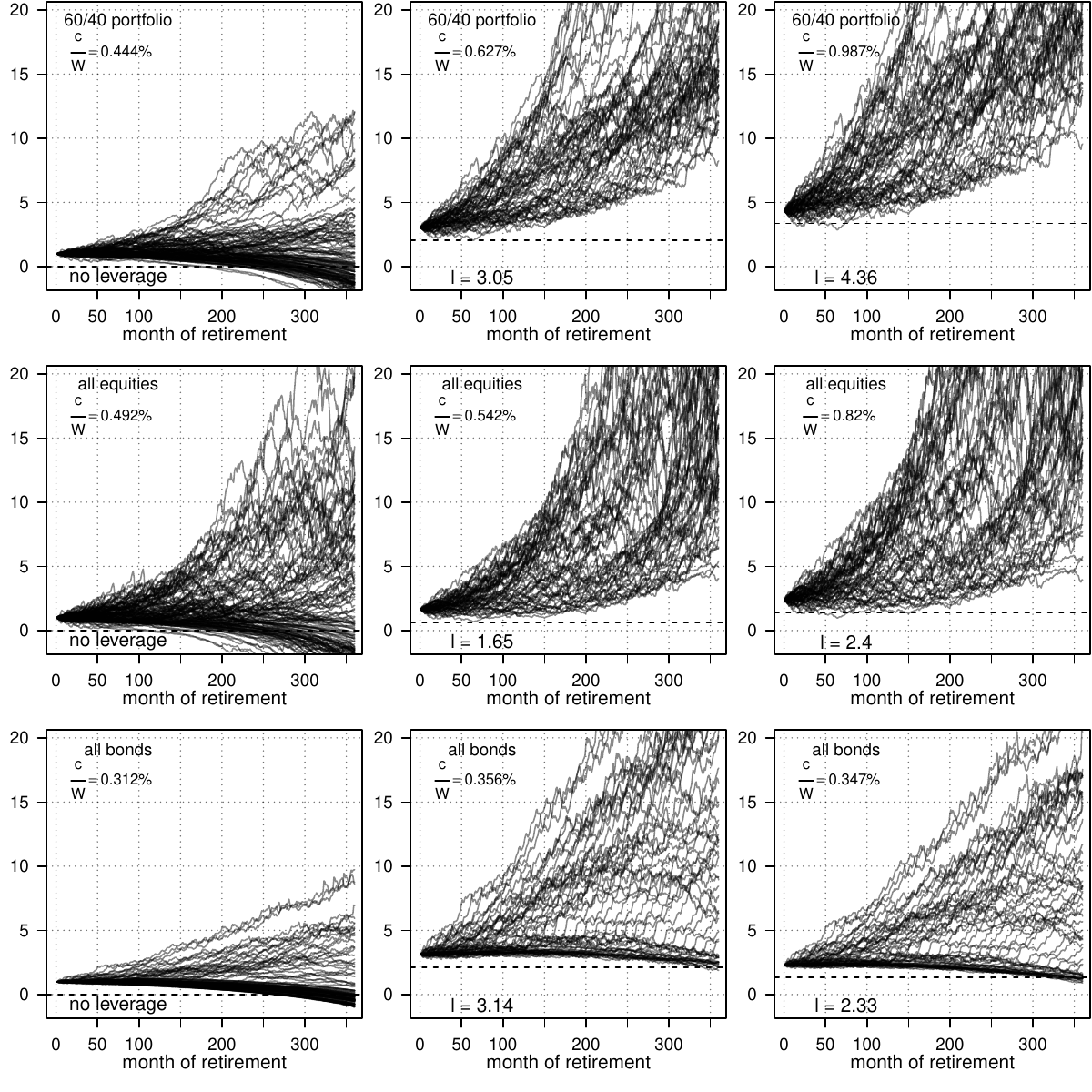}
\caption{30-year retirements, derived from historical asset returns, funded by portfolios of different compositions (rows) and leverage ratios. The (first-month) withdrawal rate, presented as $c/W$, is the same for all retirements within a panel, but differs among panels. Each jagged continuous line is the monthly trajectory of a portfolio's remaining value relative to the initial wealth $W$. Horizontal dashed lines represent a failed retirement, either bankruptcy (in the absence of leverage) or exhaustion of the initial margin (for levered portfolios).}
\label{fig:historical}
\end{figure*}

I subject each set of results in table \ref{ta:leveredportfolios} to annual historical simulations, presented in figure \ref{fig:historical}.
For each set of un-levered results I simulate 30-year retirements beginning in January 1871, in January 1872, and so on, in 12-month increments to January 1995; for the levered results I simulate 30-year retirements beginning in January 1934 (since that is when my price-of-leverage time series begin), in January 1935, and so on, again in 12-month increments to January 1995.
All simulated retirements use a time step of one calendar month, and assume monthly rebalancing of portfolios and month-on-month consumption growth $s$ of 0.3\%.

In the classic case of an un-levered 60/40 portfolio (top-left panel of figure \ref{fig:historical}), the second-order formulae suggest withdrawing and consuming 0.444\% of one's initial wealth $W$ in the first month of retirement.
On average this works well: across the 125 simulated retirements the mean wealth remaining after 30 years is 97\% of $W$ and the median is $-12$\%.
The median retirement modestly fails --- and 48\% of simulated retirements deplete the portfolio before the planned end of retirement --- but the formulae get as close as one could reasonably expect to the goal of spending precisely 100\% of $W$ in precisely 30 years.
With returns fluctuating over time, no rigid rule based on spending a pre-calculated percentage of $W$ in the first month, and mechanically increasing that percentage at a fixed relative rate, could fit every retirement.

Introducing leverage (as in the top-centre and top-right panels of fig.\ \ref{fig:historical}) enables a higher rate of consumption while reducing the risk of outright bankruptcy.
Retirements funded by a levered 60/40 portfolio could have supported more than twice the un-levered rate of consumption, with no outright bankruptcies.
Importantly, that cheerful finding is subject to two vital caveats.
The first is that, being built on levered portfolios, these simulated retirements cover only the 1934--2025 period, when stocks performed relatively well.
The other caveat is that when leverage is in use, the criterion for a retirement failure is not outright bankruptcy (the value of the portfolio prematurely reaching zero) but a decrease in the portfolio's value equal to the retiree's initial wealth, which would typically trigger a margin call (a demand from the leverage-supplying brokerage that the retiree somehow add liquidity to their portfolio) or forced liquidation of some of the portfolio.
That latter kind of failure does occur in 2\% of the lower-leverage simulations and in 3\% of the higher-leverage simulations.

At the same time, those failure rates are low, and perhaps unexpectedly low given that my derivations of the optimal-leverage formulae do not include the concepts of drawdown limits or margin calls.
The formulae nonetheless recommend leverage ratios that prove relatively conservative; margin calls would have been uncommon, and the median levered-60/40 trajectory does not come close to breaching the margin-exhaustion lower bound.
Better still, when these levered portfolios do breach that bound, they do so early and temporarily.
While un-levered 60/40 portfolios that fail often do so irreversibly and in the middle of the intended retirement, when a retiree might be older and less able to secure a backup income to rescue their portfolio, levered 60/40 portfolios that fail do so temporarily and in the first 6 years or so of retirement, when a retiree might be relatively young and better able to rescue the portfolio with small, temporary injections of liquidity.

Moving from 60/40 portfolios to all-equities portfolios, the formulae unsurprisingly suggest a higher consumption rate ($c/W = 0.492\%$ instead of $c/W = 0.444\%$) in the absence of leverage.
Despite this higher consumption rate the median retirement ends at 48\% of its initial $W$, although, again, that the formulae work on average does not mean that every retirement would have succeeded: 46\% of the simulated retirement trajectories fall below zero.

Like the 60/40 portfolio, levering the all-equities portfolio would have allowed notably higher consumption, and improved returns would have more than compensated for that higher consumption.
None of the $l = 1.65$ trajectories, and only 5\% of the $l = 2.40$ trajectories, exhaust the portfolio's initial margin.

Finally, the all-bonds portfolios.
Without leverage, the second-order formulae suggest a first-month $c/W$ of only 0.312\%, leading to bankruptcy in 65\% of simulated histories.
That is high enough (notably higher than the 50\% one might expect) to give a little pause.
Similarly, the failure rates with leverage are higher than the failure rates for levered majority-equities portfolios.
Interestingly, however, when the all-bonds portfolios are levered, the retirement failures all cluster near the end of retirement, 318--353 months into the intended 360-month retirement, an encouraging outcome for formulae built on the assumption of retirement wealth being completely spent down at the end of retirement.
The failure rates, moreover, are again low under leverage: only 5\% when $l = 3.14$ and 19\% when $l = 2.33$.

Across the 3 types of portfolio, the formulae produce roughly the expected result without leverage; about half of simulated retirements undershoot (suffer total portfolio depletion in under 30 years) and about half overshoot (enjoy a positive portfolio value for 30 years).
Specifically, 48\% for 60/40 portfolios, 46\% of all-equities portfolios, and 65\% of all-bonds portfolios run out of wealth before the intended 30 years.
With second-order-optimal leverage, the failure rate decreases to just 19\% for all-bonds portfolios at a leverage ratio of 233\%, and to 5\% or less for every other portfolio and leverage ratio backtested, despite higher withdrawal rates making stronger demands of the levered portfolios.
Indeed, the four levered majority-equities portfolios simulated here all delivered withdrawal rates $> 6.6\%$ per year alongside failure rates of 5\% or less, indicating that leveraging retirement wealth could improve on the 4\% rule of thumb.

\section{Conclusion}

From a simple discrete-time mathematical model of retirement consumption funded by a portfolio, I have derived approximate but illustrative closed-form expressions for how aggressively a retiree can withdraw from their portfolio.
For this purpose the portfolio can be summarized by a single parameter $\gamma$, which is a function of the retiree's rate of consumption growth $s$ and the lowest-order moments of the portfolio's probability distribution of returns.
The variance of returns and $s$ are interchangeable in the second-order formula for $\gamma$, indicating that, to second order, variance in returns and consumption growth impose the same drag on $\gamma$ and the portfolio's ability to fund a retirement.

The retiree's feasible rate of withdrawal is an explicit function of $\gamma$ and the duration of retirement $t$, allowing ready estimation.
Fed with historical data about returns to US equities and bonds, my mathematical results suggest why Bengen found 4\% to be a safe withdrawal rate for portfolios combining US large-cap stocks and US bonds, as opposed to, say, 2\% or 6\% or simply the real rate of return to stocks.
The safe withdrawal rate is that low because retirement portfolios seldom use leverage, and so have their expected returns dragged down by including bonds; because consumption growth, variance (and kurtosis) of returns, and their interaction, reduce the effective rate of return on which a retiree or other dissaver can rely; and because the stereotypical retiree wishes to retire for the rest of their life, however long it is, introducing longevity risk that obliges a retiree to be concerned with more than the 30 years Bengen suggested as a benchmark duration.

I also analyzed the role of leverage and derived approximations for the optimal leverage of retirement portfolios.
Despite exacerbating volatility, leveraging equities-based retirement portfolios appears to be optimal.
That leverage increases the variance of returns more than their mean explains why retirees ought not lever up their portfolios as aggressively as possible, but does not mean that leverage should be forsworn.
Were leverage available to retirees at the federal funds rate plus a one-point-per-year spread, leverage ratios of 1.6--3.1 might have been optimal for portfolios that were mostly S\&P 500 stocks.
Historical simulations indicate that despite amplifying volatility, sufficiently cheap leverage can reduce the risk of a retiree depleting all of their wealth prematurely, and could have rendered withdrawal rates of 6.6\% per year or more reasonably safe.


\begin{thebibliography}{6}

\bibitem{Bengen94}
William P Bengen (1994).
Determining withdrawal rates using historical data.
\textit{Journal of Financial Planning}, \textbf{7}, 171--180.

\bibitem{DeNardi16}
Mariacristina De Nardi, Eric French, John Bailey Jones (2016).
Savings After Retirement: A Survey.
\textit{Annual Review of Economics}, \textbf{8}, 177--204.

\bibitem{FREDCPI}
Federal Reserve Economic Data (2024).
Consumer Price Index: All Items: Total for United States.
Time series \texttt{CPALTT01USM657N}, retrieved from \url{https://fred.stlouisfed.org/series/CPALTT01USM657N} on 24 April 2024.

\bibitem{Bontrager23}
Marvin Bontrager (2023).
X/Twitter post posted 9 November 2023, and retrieved from \url{https://x.com/mbontrager5/status/1722478848573329702} on 15 November 2025.

\bibitem{Ramsey23}
The Ramsey Show (2023).
You Can’t Win With Money if You Don’t Know Where Your Money Is | November 2, 2023.
YouTube video streamed live on 2 November 2023, retrieved from \url{https://www.youtube.com/watch?v=Xg4Z8EQY3Ao&t=4430s} on 15 November 2025.

\bibitem{FREDTBill}
Federal Reserve Economic Data (2025).
3-Month Treasury Bill Secondary Market Rate, Discount Basis.
Time series \texttt{TB3MS}, retrieved from \url{https://fred.stlouisfed.org/series/TB3MS} on 11 July 2025.

\bibitem{FREDFEDFUNDS}
Federal Reserve Economic Data (2025).
Federal Funds Effective Rate.
Time series \texttt{FEDFUNDS}, retrieved from \url{https://fred.stlouisfed.org/series/FEDFUNDS} on 25 July 2025.

\bibitem{Asness96}
Clifford S Asness (1996).
Why \emph{Not} 100\% Equities.
\textit{Journal of Portfolio Management}, \textbf{22}(2), 29--34.

\end{thebibliography}
\end{document}